\documentclass[doublecol]{epl2}
\usepackage{amssymb}
\usepackage{mathrsfs}
\usepackage{amsmath}
\usepackage{flushend}

\title{Quenched mean-field theory for the majority-vote model on complex networks}
\shorttitle{Quenched mean-field theory for the majority-vote model on complex networks} 

\author{Feng Huang\inst{1} \and Hanshuang Chen\inst{2}\thanks{ \email{chenhshf@ahu.edu.cn}} \and Chuansheng Shen\inst{3}}
\shortauthor{F. Huang, H. Chen and C. Shen}

\institute{
  \inst{1} School of Mathematics and Physics, Anhui Jianzhu University, Hefei 230601, China \\
  \inst{2} School of Physics and Material Science, Anhui University, Hefei, 230601, China \\
\inst{3} Department of Physics, Anqing Normal University, Anqing,
246011, China}


\abstract{The majority-vote (MV) model is one of the simplest
nonequilibrium Ising-like model that exhibits a continuous
order-disorder phase transition at a critical noise. In this paper,
we present a quenched mean-field theory for the dynamics of the MV
model on networks. We analytically derive the critical noise on
arbitrary quenched unweighted networks, which is determined by the
largest eigenvalue of a modified network adjacency matrix. By
performing extensive Monte Carlo simulations on synthetic and real
networks, we find that the performance of the quenched mean-field
theory is superior to a heterogeneous mean-field theory proposed in
a previous paper [Chen \emph{et al.}, Phys. Rev. E 91, 022816
(2015)], especially for directed networks.}

\begin{document}

\maketitle

\section{Introduction}
Dynamical processes taking place on complex networks are often used
to model a wide variety of phenomena
\cite{PRP06000175,PRP08000093,boccaletti2014structure,PRP2016,boccaletti2016explosive}.
Examples include spreading of diseases or opinions through a
population \cite{RevModPhys.87.925,wang2016statistical}, neural
activity in the brain \cite{PNAS2009}, and cascading failure on
power grid \cite{Nature2010}. Owing to the inherent randomness and
heterogeneity in the interacting patterns, it has lead to dynamics
on complex networks drastically different from those on regular
lattices in Euclidean space \cite{RMP08001275}. So far, unveiling
the relationship between the topologies of the underlying networks
and the dynamics on them is still a hot topic of considerable
attention.

Mean-field theory (MFT) is the most commonly used to obtain
(relatively) low-dimensional descriptions for the systems under
study. For homogeneous networks, MFT is quite accurate. However, for
heterogeneous networks such as scale-free networks (SFN), ordinary
MFT fails since degrees of nodes are largely different from each
other. An improved MFT, termed heterogeneous MFT (HMFT) was proposed
for the analysis of general dynamical processes on heterogeneous
networks. By assuming that nodes with the same degree are
statistically equivalent, the HMFT writes down a set of degree-based
equations for governing the evolutions of dynamical processes on
complex networks \cite{RevModPhys.87.925,RMP08001275}. Some
well-known conclusions of the HMFT include the absence of epidemic
threshold in susceptible-infected-susceptible (SIS) model
\cite{PRL01003200,PhysRevLett.90.028701} and the anomalous behavior
of Ising model
\cite{PHA02000260,PLA02000166,PRE02016104,EPB02000191,PhysRevLett.104.218701}
in SFN with degree exponent less than 3. It is widely accepted that
the HMF theory is exact only when the underlying networks are
annealed where dynamical correlations among neighboring nodes are
absent \cite{RMP08001275,RevModPhys.87.925}. An annealed network
requires that the time scale of network evolution is much faster
than that of dynamics on it. In other words, an annealed network is
far from static, in contrasts with a network where links are fixed
permanently in time. While the latter network is referred to as a
quenched network. Almost concurrently, a quenched MFT (QMFT) was
proposed to study the dynamics on arbitrary quenched networks. The
basic idea of QMFT is to derive evolution equations for each node by
assuming that the dynamic state of each node is statistically
independent of the state of its nearest neighbors. For example, the
QMFT predicts that the epidemic threshold of SIS model equals to the
inverse of the largest eigenvalue of the adjacency matrix of the
underlying network \cite{wang2003epidemic,Mieghem2009,Gomez2010}.
Furthermore, there are several extensions of HMFT-based
\cite{Eames2002,PhysRevLett.107.068701,PhysRevLett.111.068701,Mata2014,PhysRevLett.115.078701,PhysRevLett.116.258301}
and QMFT-based \cite{PhysRevE.85.056111,Mata2103,kiss2015exact}
theories have been developed that take into account the role of
dynamical correlations.

The majority-vote (MV) model is a simple nonequilibrium Ising-like
system with up-down symmetry \cite{JSP1992}. In the model, each
individual is assigned to a binary spin variable $\pm 1$. In each
time step, each spin tends to align with the local neighborhood
majority but with a noise parameter $f$ giving the probability of
misalignment. As $f$ increases, the model presents a continuous
order-disorder phase transition at a critical value $f_c$. The MV
model has been extensively studied for various interacting
substrates, such as regular lattices
\cite{PhysRevE.75.061110,PhysRevE.81.011133,PhysRevE.89.052109,PhysRevE.86.041123},
random graphs \cite{PhysRevE.71.016123,PA2008}, small-world networks
\cite{PhysRevE.67.026104,IJMPC2007,PA2015}, scale-free networks
\cite{IJMPC2006(1),IJMPC2006(2)}, and modular networks
\cite{CPL2015}.

In a recent work \cite{PhysRevE.91.022816}, we have used the HMFT to
derive the critical noise $f_c$, which is determined by
\begin{eqnarray}
(1 - 2{f_c})\sum\limits_k {\frac{{{k^2}P(k)}}{{\left\langle k
\right\rangle }}} {2^{1 - k}}C_{k - 1}^{\left\lceil {(k - 1)/2}
\right\rceil } = 1, \label{eq1}
\end{eqnarray}
where $P(k)$ is degree distribution defined as the probability that
a node chosen at random has degree $k$, and $\left\langle k
\right\rangle$ is the average degree. $C_{k}^n={k!}/[n!(k-n)!]$ is
the binomial coefficient and $\left\lceil \cdot \right\rceil$ is the
ceiling function. For large $k$, by Stirling's approximation, $C_{k
- 1}^{\left\lceil {(k-1)/2} \right\rceil } \approx {2^{k - 1}}/\sqrt
{k \pi /2}$, and $f_c$ is thus given explicitly,
\begin{eqnarray}
{f_c} = \frac{1}{2} - \frac{1}{2}\sqrt {\frac{\pi }{2}}
\frac{{\left\langle k \right\rangle }}{{\left\langle {{k^{3/2}}}
\right\rangle }}, \label{eq2}
\end{eqnarray}
where $\left\langle {{k^n}} \right\rangle  = \sum\nolimits_k {{k^n}}
P(k)$ is the $n$th moment of degree distribution.

As mentioned before, HMFT is only exact for annealed networks. While
for quenched networks, QMFT is expected to be more accurate than
HMFT. To fill this gap, here we develop a QMFT for the dynamics of
the MV model. We show that the critical noise can be determined by
the leading eigenvalue of a modified adjacency matrix of the
underlying network. By performing extensive Monte Carlo (MC)
simulations on diverse types of networks, we find that the QMFT is
superior to the HMFT.

\section{Model} \label{sec2}
Let us first define the MV model on an unweighed network with size
$N$, where the network is described by an adjacency matrix
$\textbf{A}$. The elements of $\textbf{A}$ are defined as $A_{ij}=1$
if there exists an edge from node $i$ to node $j$, and $A_{ij}=0$
otherwise. The spin of each node is assigned to a binary variable
$\sigma_i \in \{-1, +1\}$ $(i = 1, \ldots ,N)$. The system evolves
as follows: for each node $i$, we determine the majority spin of its
neighborhood. With probability $f$ the node $i$ takes the opposite
sign of the majority, otherwise it takes the same spin as the
majority. $f$ is called the noise parameter. In this way, the spin
flipping probability of node $i$ is given by
\begin{eqnarray}
w_i = \frac{1}{2}\left[ {1 - \left( {1 - 2f} \right){\sigma
_i}S\left( \Theta_i \right)} \right], \label{eq3}
\end{eqnarray}
with
\begin{eqnarray}
\Theta_i = {\sum\limits_j {{A_{ij}}{\sigma _j}} }, \label{eq3s}
\end{eqnarray}
where $S(x)=sgn(x)$ if $x\neq0$ and $S(0)=0$. In the latter case the
spin $\sigma_i$ is flipped to $\pm 1$ with equal probabilities ${1
\mathord{\left/ {\vphantom {1 2}} \right. \kern-\nulldelimiterspace}
2}$. For $f=0$, the MV model is equivalent to the zero-temperature
Ising model with Glauber dynamics \cite{PNAS2005,JSM2006}.

The MV model does not only play an important role in the study of
nonequilibrium phase transitions, but also helps to understand
opinion dynamics in social networks \cite{RMP09000591}. In this
model, binary spins can represent two opposite opinions, and the
noise parameter $f$ plays the role of the temperature in equilibrium
systems.

\section{Results}  \label{sec3}

To proceed the QMFT, let us define $p_i$ as the probability that the
spin of node $i$ takes $\sigma_i=1$. The dynamical evolution of
$p_i$ is governed by the rate equation,
\begin{eqnarray}
\frac{{d{p_i}}}{{dt}} =  - {p_i}w_i^ +  + \left( {1 - {p_i}}
\right)w_i^ -, \label{eq4}
\end{eqnarray}
where $w_i^+$ ($w_i^-$) is the flipping probabilities of node $i$
with spin $+1$ ($-1$). According to the dynamics of the MV model,
$w_i^+$ can be written as the sum of two parts,
\begin{eqnarray}
w_i^ +  = f{\varphi _i} + \left( {1 - f} \right)\left( {1 - {\varphi
_i}} \right), \label{eq5}
\end{eqnarray}
where the first part is the product of the probability $\varphi_i$
of the majority spin among the neighborhood of the node $i$ being
$+1$ and the probability $f$ of the minority rule being applied, and
the second part is the product of the probability $1-\varphi_i$ of
the minority spin among the neighborhood of the node $i$ being $+1$
and the probability $1-f$ of the majority rule being applied.
Likewise, $w_i^-$ is expressed as,
\begin{eqnarray}
w_i^ -  = \left( {1 - f} \right){\varphi _i} + f\left( {1 - {\varphi
_i}} \right). \label{eq6}
\end{eqnarray}
Considering $w_i^+ + w_i^-=1$, Eq. (\ref{eq4}) reduces to
\begin{eqnarray}
\frac{{d{p_i}}}{{dt}} =  - {p_i} + w_i^ -. \label{eq7}
\end{eqnarray}
The probability $\varphi_i$ can be written as
\begin{eqnarray}
{\varphi _i} = \sum\limits_{n = \left\lceil {k_i^{out}/2}
\right\rceil }^{k_i^{out}} {\left(1 - \frac{1}{2}{\delta
_{n,k_i^{out}/2}}\right)} \xi_{n,k_i^{out}},\label{eq8}
\end{eqnarray}
where $k_i^{out}$ is the outdegree of node $i$, and $\delta$ is the
Kronecker delta function. $\xi_{n,k_i^{out}}(i)$ is the probability
that there are $n$ up spins among the neighborhood $\mathcal {N}(i)$
of node $i$, which can be calculated by
\begin{eqnarray}
\xi_{n,k_i^{out}} = {\sum\limits_{\mathcal {U}(i) \subseteq \mathcal
{N}(i)} {\prod\limits_{j \in \mathcal {U}(i)} {{p_j}} } }
\prod\limits_{j \in  \mathcal {\bar U}(i) }{(1 - {p_j})},\label{eq9}
\end{eqnarray}
where $\mathcal {U}(i)$ are all the subsets of $\mathcal {N}(i)$
that contains $n$ neighbor(s) of node $i$, $\left|{\mathcal
{U}(i)}\right|=n\in\ [0,k_i^{out}]$. $\mathcal {\bar U}(i)$ is the
complement of the subset $\mathcal {U}(i)$, i.e., $\mathcal
{U}(i)\cup\mathcal {\bar U}(i)=\mathcal {N}(i)$ and $\mathcal {U}(i)
\cap \mathcal {\bar U}(i)=\emptyset$. Obviously, there are
$C_{k_i^{out}}^n$ possibilities for the subsets $\mathcal {U}(i)$.

Since $\varphi _i=1/2$ at $p_i=1/2$ for all $i$, one can easily
check that $p_i=1/2$ is always a stationary solution of Eq.
(\ref{eq7}). For convenience, let us note this trivial solution as
$\textbf{p}^*$. Such a trivial solution corresponds to a disordered
phase. An ordered phase can emerge when the noise parameter $f$ is
lower than the so-called critical value, $f_c$, below which the
trivial solution $p_i=1/2$ loses its stability. Thus, $f_c$ is
determined by the maximal eigenvalue of Jacobian matrix that is
null, ${\Lambda _{\max }}\left( \textbf{J} \right) = 0$, where the
elements of the matrix $\textbf{J}$ are given by,
\begin{eqnarray}
{J_{ij}} =  - {\delta _{ij}} + {\left. {\left( {1 - 2f}
\right)\frac{{\partial {\varphi _i}}}{{\partial {p_j}}}}
\right|_{\textbf{p}^*}}.\label{eq10}
\end{eqnarray}
If $j \notin \mathcal {N}(i)$, $\varphi_i$ is considered to be
independent of $p_j$, so that ${\frac{{\partial {\varphi
_i}}}{{\partial {p_j}}}}= 0$. If $j \in \mathcal {N}(i)$,
${\frac{{\partial {\varphi _i}}}{{\partial {p_j}}}}$ at
$\textbf{p}^*$ can be derived as follows. For $j \in \mathcal
{U}(i)$ and $\left|{\mathcal {U}(i)}\right|=n$, the remaining $n-1$
nodes different from node $j$ are included in the set $\mathcal
{U}(i)$. There exist $C_{k_i^{out} - 1}^{n - 1}$ possibilities for
$\mathcal {U}(i)$, and its contribution to the partial derivation
$\partial {\varphi_i}/\partial {p_j}$ is positive in terms of Eq.
(\ref{eq9}). For $ j \in \mathcal {\bar U}(i)$, there exist
$C_{k_i^{out} -1}^{n }$ possibilities for $\mathcal {U}(i)$, and its
contribution to the partial derivation is negative. $C_{k_i^{out}
}^{n}=C_{k_i^{out} -1}^{n }+C_{k_i^{out} - 1}^{n - 1}$ holds for any
$n$. Therefore, we have
\begin{eqnarray}
{\left. {\frac{{\partial {\varphi _i}}}{{\partial {p_j}}}}
\right|_{\textbf{p}^*}} &=& A_{ij} \sum\limits_{n = \left\lceil
{k_i^{out}/2} \right\rceil }^{k_i^{out}}  {\left( {1 - \frac{1}{2}
{\delta _{n,k_i^{out}/2}}} \right)} \nonumber \\ & \times & {\left(
{C_{k_i^{out} - 1}^{n - 1} - C_{k_i^{out} - 1}^n}
\right)} {\left( {\frac{1}{2}} \right)^{k_i^{out} - 1}} \nonumber \\
&=& {A_{ij}}C_{k_i^{out} - 1}^{\left\lceil {k_i^{out}/2}
\right\rceil - 1}{2^{1-k_i^{out}}}.\label{eq11}
\end{eqnarray}
Substituting Eq. (\ref{eq11}) into Eq. (\ref{eq10}), we obtain
\begin{eqnarray}
{J_{ij}} =  - {\delta _{ij}} + \left( {1 - 2f}
\right){A_{ij}}C_{k_i^{out} - 1}^{\left\lceil {k_i^{out}/2}
\right\rceil  - 1}{2^{1-k_i^{out} }}. \label{eq12}
\end{eqnarray}
Again, $f_c$ is obtained when the largest eigenvalue of $\textbf{J}$
is null, yielding
\begin{eqnarray}
f_c= \frac{1}{2} \left(1- \frac{1}{\Lambda_{max}( {{\tilde
{\textbf{A}}}} )}\right),\label{eq13}
\end{eqnarray}
where the elements of the matrix $\tilde {\textbf{A}}$ are
\begin{eqnarray}
{{\tilde A}_{ij}} = {A_{ij}}C_{k_i^{out} - 1}^{\left\lceil
{k_i^{out}/2} \right\rceil  - 1}{2^{1 - k_i^{out}}}.\label{eq14}
\end{eqnarray}
Eq. (\ref{eq13}) is the central result of the present work. From the
Perron-Frobenius theorem, since $\tilde A_{ij}$ is non-negative, and
assuming that it is irreducible, its largest eigenvalue is real and
positive.

For $f<f_c$, analytically deriving nontrivial solution of $p_i$ is
generally impossible. To this end, one can numerically iterate
$N$-intertwined Eq. (\ref{eq7}) at stationary, $p_i=\varphi_i$.
However, direct calculation of $\varphi_i$ is not practical, since
the combination number of the subsets $\mathcal{U}(i)$ is
tremendously large, especially for nodes with high outdegrees. To
overcome this difficulty, we adopt a recursive formulae developed in
Ref.\cite{Kuo2003}. The probability $\xi_{n,k_i^{out}}$ in Eq.
(\ref{eq9}) can be calculated by
\begin{eqnarray}
\xi_{k,j}=(1-p_j) \xi_{k,j-1}+ p_j \xi_{k-1,j-1}, 0\leq k \leq j
\leq k_i^{out}. \label{eq15}
\end{eqnarray}
The boundary conditions for Eq.(\ref{eq15}) are
$\xi_{-1,j}=\xi_{j+1,j}=0$ ($j=0,\cdots,k_i^{out}$), and
$\xi_{0,0}=1$. Once $p_i$ is obtained, average magnetization of node
$i$ can be calculated as $m_i=2 p_i-1$, and average magnetization
per node as $m=\sum\nolimits_{i = 1}^N{m_i}/N$.

To numerically determine the critical noise $f_c$ from MC
simulations, we calculate the magnetization $m$ and the
susceptibility $\chi$, defined as
\begin{eqnarray}
m = {{{\left\langle {\frac{1}{N}\sum\limits_{i = 1}^N {{\sigma _i}}
} \right\rangle }}},
\end{eqnarray}
\begin{eqnarray}
\chi  = N\left( \left\langle {m^2} \right\rangle  - {\left\langle m
\right\rangle}^2 \right),
\end{eqnarray}
where ${\left\langle  \cdots  \right\rangle }$ denotes time averages
taken in the stationary regime.

Let us now start the validation of the QMFT for a simple homogeneous
network, the random $k$-regular networks (RkRN). For RkRN, all nodes
have the same degree $k$ and degree distribution follows $P(k) =
\delta(\left\langle k \right\rangle)$, while the edges among nodes
are linked at random. From ${\Lambda _{\max }}(\mathbb{\tilde A}) =
C_{k - 1}^{\left\lceil {k/2} \right\rceil  - 1}{2^{1 -
k}}\Lambda_{\max}(\mathbb{A})$ and $\Lambda_{\max}(\mathbb{A})=k$,
we arrive at the critical noise of RkRN,
\begin{eqnarray}
f_c^{RkRN} = \frac{1}{2} - \frac{1}{{kC_{k - 1}^{\left\lceil {k/2}
\right\rceil  - 1}{2^{2 - k}}}}.\label{eq16}
\end{eqnarray}
For large $k$, $C_{k - 1}^{\left\lceil {k/2} \right\rceil  - 1}
\approx {2^{k - 1}}/\sqrt {(k - 1)\pi /2}$ by Stirling's
approximation, and thus
\begin{eqnarray}
f_c^{RkRN} \approx \frac{1}{2} - \frac{1}{{2k}}\sqrt {\frac{{(k -
1)\pi }}{2}}  \approx \frac{1}{2} - \frac{1}{2}\sqrt {\frac{\pi
}{{2k}}},\label{eq17}
\end{eqnarray}
which is consistent with the result of HMFT, Eq. (\ref{eq2}).
However, for any randomly heterogeneous networks, on the one hand,
it is generally impossible to analytically obtain
$\Lambda_{\max}(\mathbb{A})$, which can be instead calculated
numerically. It should be noted that there were some attempts to
examine the spectral properties of network adjacency matrices with
given degree distributions \cite{PhysRevE.87.012803}, which might be
a heuristic effect on the present work. On the other hand, the
result of QMFT is different from that of HMFT.

\begin{figure}
\centerline{\includegraphics*[width=1.0\columnwidth]{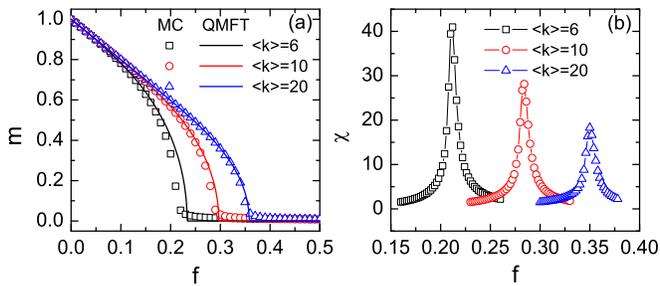}}
\caption{(Color online). (a) The magnetization $m$ and (b) the
susceptibility $\chi$ as a function of the noise intensity $f$ on
RkRN with network size $N=10000$. The lines and symbols in (a)
indicate the QMFT and simulation results, respectively.
\label{fig1}}
\end{figure}

Fig. \ref{fig1}(a) shows the magnetization $m$ as a function of
noise intensity $f$ in RKRN with $N=10000$ and three different
average degree $\left\langle k \right\rangle=6$, $10$, and $20$. The
networks are generated according to the Molloy-Reed model
\cite{RSA95000161}: each node is assigned a random number of stubs
$k$ that is drawn from a given degree distribution. Pairs of
unlinked stubs are then randomly joined. The lines and symbols in
Fig. \ref{fig1}(a) indicate the QMFT and simulation results,
respectively. Fig. \ref{fig1}(b) shows the susceptibility $\chi$ as
a function of $f$. $\chi$ exhibits a sharp peak at the critical
noise $f_c$. MC simulation gives $f_c=0.212$, $0.284$, and $0.350$
for $\left\langle k \right\rangle=6$, $10$, and $20$, respectively.
The QMFT predicts $f_c=0.233$, $0.297$, and $0.358$ for
$\left\langle k \right\rangle=6$, $10$, and $20$, respectively. The
discrepancy between them becomes more apparent for smaller
$\left\langle k \right\rangle$. The better accuracy of the QMFT for
larger $\left\langle k \right\rangle$ is intuitive since topological
distance among nodes decreases as the average degree increases
making the mean-field premise a more credible hypothesis.

\begin{figure}
\centerline{\includegraphics*[width=1.0\columnwidth]{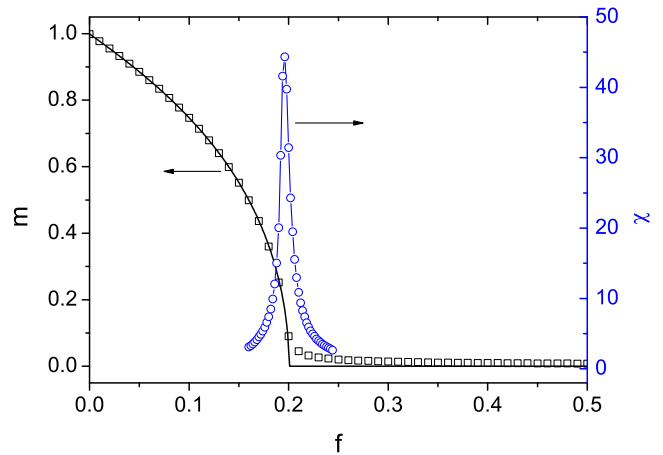}}
\caption{(Color online). The magnetization $m$ (left axis) and the
susceptibility $\chi$ (right axis) as a function of the noise
intensity $f$ on a directed network with network size $N=10000$ and
$\left\langle k^{out} \right\rangle\equiv\left\langle k^{in}
\right\rangle=5$, generated by a RkRN. The lines and symbols (left
axis) indicate the QMFT and simulation results, respectively.
\label{fig2}}
\end{figure}

We next demonstrate the results on directed networks. A directed
network is constructed as follows. We start with an undirected
network, and then turn each undirected edge, $i\leftrightarrow j$,
into a directed edge, $i\rightarrow j$ or $i\leftarrow j$ with equal
probabilities $1/2$. The average outdegree and the average indegree
of the resulting directed network are $\left\langle k^{out}
\right\rangle\equiv\left\langle k^{in} \right\rangle=\left\langle k
\right\rangle$/2, where $\left\langle k \right\rangle$ is the
average degree of the original undirected network. Unlike HMFT, the
QMFT does not require that the underlying networks are undirected.
Fig. \ref{fig2} shows $m$ (left axis) and $\chi$ (right axis) as a
function of $f$ on a directed network, generated by a RkRN with
$N=10000$ and $\left\langle k\right\rangle=10$ (named as D-RkRN for
shorts). MC simulation and QMFT show the critical noises are
respective $f_c=0.196$ and $0.200$, as shown by the symbols and
lines (left axis). There are in excellent agreement between them.
Moreover, we substitute $k$ in Eq. (\ref{eq1}) and Eq. (\ref{eq2})
with $k^{out}$, which enables HMFT to be applied to directed
networks. It is found that the HMFT result is $f_c=0.230$ (see Table
\ref{tab1} for comparison). As expected, QMFT is obviously superior
to HMFT in predicting critical point on directed networks.

To further test the potential of our QMFT on other networks, we
consider synthetic networks: Erd\"os-R\'enyi (ER) random networks
(undirected) \cite{ER1960} Barab\'asi-Albert (BA) scale-free
networks (undirected) \cite{Science.286.509}, directed ER (D-ER)
networks and directed BA (D-BA) networks, as well as real networks:
the Email network of the University at Rovira i Virgili (containing
1133 nodes and 5451 directed edges) \cite{PhysRevE.68.065103} and
Wikivote network (containing 7115 nodes and 103689 directed edges)
\cite{leskovec2010signed}. The sizes and the average degrees of the
ER network and BA network we use are both $N=10000$ and
$\left\langle k \right\rangle=6$. The results for the critical noise
are summarized in Table \ref{tab1}. One can see that the
performances of the QMFT on all the listed networks are better than
HMFT, especially for directed networks. The reason why HMFT does not
work well in directed networks maybe lie in an assumption of degree
uncorrelated networks used in HMFT to obtain explicitly the
expression of $f_c$. In general, such an assumption is no longer
valid for directed networks. While for QMFT, it is only based on the
network adjacency matrix that attains the full information of the
network topology, whether the network is undirected or directed.

\begin{table}[h]
\centering \caption{Comparison of critical noise $f_c$ on seven
networks.} \label{tab1}
\begin{tabular*}{8cm}{@{\extracolsep{\fill}}llllllll}
\hline\hline Network & Type &  MC & QMFT & HMFT  \\
\hline
  D-RkRN & directed & 0.196 & 0.200 & 0.230 \\
  ER & undirected & 0.272 & 0.310 & 0.321 \\
  BA & undirected & 0.236 & 0.261 & 0.265 \\
  D-ER & directed & 0.116 & 0.121 & 0.180 \\
  D-BA & directed & 0.190 & 0.207 & 0.259 \\
  Email & directed & 0.073 & 0.102 & 0.268 \\
  Wikivote & directed & 0.312 & 0.366 & 0.437 \\

        \hline \hline

\end{tabular*}\\
\end{table}

Recently, an inertial effect was added to the dynamics of the MV
model, such that the spin-flip probability of each node depends not
only on the its neighboring spins, but also on its own spin
\cite{PhysRevE.95.042304}. In the inertial MV model, Eq. (\ref{eq4})
is rewritten as
\begin{eqnarray}
\Theta_i = (1-\theta){\sum\nolimits_j {{A_{ij}}{\sigma _j}}
}/k_i^{out}+\theta \sigma_i, \label{eq21}
\end{eqnarray}
where $\theta  \in \left( {0,0.5} \right]$ controls the strength of
the inertia. Note that for $\theta=0$ one recovers the original MV
model.

Interestingly, the order-disorder phase transition is changed from a
usual continuous or second-order type to a discontinuous or
first-order one when $\theta$ is larger than a critical value
$\theta_c$ \cite{PhysRevE.95.042304,Chaos27.081102}. If the inertia
is placed to partial nodes, a mixture of discontinuous and
continuous phase transitions from fully ordered phase to partially
ordered phase and then to disordered phase was observed
\cite{PhysRevE.96.042305}. To derive QMFT in the inertial MV model,
the probability $\varphi_i$ in Eq. (\ref{eq6}) and Eq. (\ref{eq7})
should be replaced by $\varphi_i^+$ and $\varphi_i^-$, respectively.
Here $\varphi_i^+$ ($\varphi_i^-$) is the conditional probability
that the majority spin among the neighborhood of node $i$ being
$+1$, providing that the spin of node $i$ is up (down). They can be
written as,
\begin{eqnarray}
{\varphi _i^\pm} = \sum\limits_{n = n_i^\pm}^{k_i^{out}} {\left(1 -
\frac{1}{2}{\delta _{n,n_i^\pm}}\right)}
\xi_{n,k_i^{out}},\label{eq22}
\end{eqnarray}
where $n_i^+={k_i^{out}}(1-2\theta)/[2(1-\theta)]$ and
$n_i^-={k_i^{out}}/[2(1-\theta)]$ satisfying $\Theta_i = 0$ in Eq.
(\ref{eq21}). In the steady state, one has that
\begin{eqnarray}
p_i= \frac{w_i^ -}{w_i^ +  + w_i^ -}. \label{eq23}
\end{eqnarray}

\begin{figure}
\centerline{\includegraphics*[width=1.0\columnwidth]{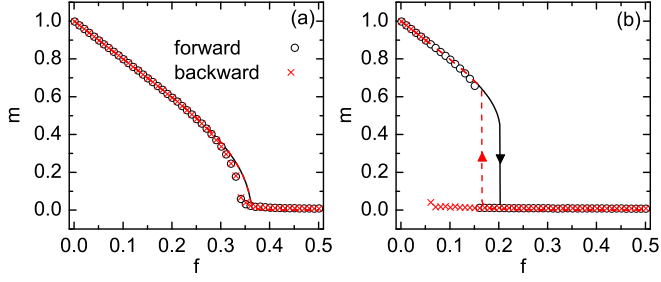}}
\caption{(Color online). The magnetization $m$ as a function of the
noise intensity $f$ on a ER network with $N=10^4$ and $\left\langle
k \right\rangle =20$ for $\theta=0.1$ (a) and $\theta=0.3$ (b). The
lines and symbols indicate the QMFT and simulation results,
respectively.} \label{fig3}
\end{figure}

Fig. \ref{fig3} shows $m$ as a function of $f$ for two different
values of $\theta$ on a ER network with $N=10^4$ and $\left\langle k
\right\rangle =20$. The simulation results (symbols) are obtained by
performing forward and backward simulations, respectively. The
former is done by calculating the stationary value of $m$ as $f$
increases from 0 to 0.5 in steps of 0.01 and using the final
configuration of the last simulation run as the initial condition of
the next run, while the latter is performed by decreasing $f$ from
0.5 to 0 with the same step. For $\theta=0.1$, forward and backward
simulations coincide indicating that the order-disorder transition
is continuous. For $\theta=0.3$, forward and backward simulations
form a hysteresis loop, a feature of a discontinuous phase
transition. The lines shows the theoretical results obtained by
numerically solving Eq. (\ref{eq23}). Although the theory agrees
qualitatively with MC simulation, there exist obvious disagreements
between them in evaluating the critical point of the discontinuous
phase transition.

\section{Conclusions} \label{sec4}

In conclusion, we have developed a QMFT for studying the MV model on
complex networks. The QMFT can not only predict the phase transition
behavior of the MV model, but also can analytically derive the
critical point of the transition. Our proposed theory has shown that
the critical point of the MV model is determined by the leading
eigenvalue of a modified adjacency matrix of the underlying network.
This result is similar but different from the results of QMFT in the
SIS model and in the Ising model. In the latter two models, the
critical points are directly determined by the leading eigenvalue of
the adjacency matrix. By extensive MC simulations on various
networks, we found that the QMFT is better than the HMFT in
predicting the critical point, especially for directed networks.
However, for networks with low average degrees, both of them
overestimate the critical point. Therefore, in the future it will be
desirable to develop high-order theories (such as pair
approximation) to obtain more accurate estimation of the critical
point of the networked MV model.

\acknowledgments We acknowledge the supports from the National
Natural Science Foundation of China (Grants No. 11405001, No.
11475003, No. 61473001), the Key Scientific Research Fund of Anhui
Provincial Education Department (Grants No. KJ2016A015) and ``211"
Project of Anhui University (Grants No. J01005106).

\bibliographystyle{eplbib}

\end{document}